\pdfoutput=1
\documentclass[%
 reprint,
superscriptaddress,
 amsmath,amssymb,
 aps,
 pra,
]{revtex4-2}

\usepackage{caption}
\usepackage{subcaption}
\usepackage{graphicx}
\usepackage{amsmath}
\usepackage{geometry}
\usepackage{hyperref}
\usepackage[T1]{fontenc}
\usepackage[utf8]{inputenc}
\usepackage{float}
\usepackage{csquotes}
\usepackage[version=4]{mhchem}
\usepackage{amssymb}
\usepackage{xcolor}
\usepackage{bbold}
\usepackage{dcolumn}
\usepackage{bm}
\usepackage{algorithm}
\usepackage{algpseudocode}

\newcommand{\beginsupplement}{%
        \setcounter{table}{0}
        \renewcommand{\thetable}{S\arabic{table}}%
        \setcounter{figure}{0}
        \renewcommand{\thefigure}{S\arabic{figure}}%
        \setcounter{equation}{0}
        \renewcommand{\theequation}{S\arabic{equation}}%
        \setcounter{section}{0}
        \renewcommand{\thesection}{S\Roman{section}}%
     }

\begin{document}
\preprint{}

\title{
Bridge the Gap between Classical and Quantum Neural Networks with Residual Connections
}

\author{Junxu Li}
\email{lijunxu1996@gmail.com}
\affiliation{Department of Physics, College of Science, Northeastern University, Shenyang 110819, China}
\date{\today}

\begin{abstract}
We introduce a Hybrid Quantum Residual Network (HQRN) and establish an exact functional correspondence between its state evolution and the dynamics of classical networks with residual connections.
When inputs are restricted to the computational basis, the HQRN reduces to its classical analog, enabling the direct translation of optimized classical weights into quantum unitary operations, effectively inheriting the landscape benefits of classical optimization.
Conversely, when processing general mixed states, the HQRN leverages off-diagonal quantum correlations to resolve features inaccessible to its classical analog.
We validate this framework through digit recognition and bipartite entanglement classification. 
Notably, HQRN achieves high classification accuracy even for adversarial separable states that mimic the marginal measurement statistics of entangled pairs.
Our results bridge the gap between classical and quantum residual learning, paving a scalable pathway for deep quantum architectures.
\end{abstract}

\maketitle

\section{Introduction}

Quantum machine learning (QML)\cite{schuld2015introduction, biamonte2017quantum, huang2021power, sajjan2022quantum, cerezo2022challenges} has attracted significant attention over the past decade.
This interest is driven not only by the remarkable success of classical machine learning\cite{lecun2015deep}, but also by the potential of quantum computation to achieve practical quantum advantage\cite{wu2021strong, daley2022practical, huang2022quantum}.
As noted by Scott Aaronson, "in machine learning like anywhere else, Nature will still make us work for those (quantum) speedups" \cite{aaronson2015read}.
A growing body of work supports this prospect, highlighting the potential for quantum-enhanced learning algorithms\cite{riste2017demonstration, huang2021information, anshu2021sample, lewis2025quantum, zhao2025entanglement}.

Despite these advances, a fundamental question remains, how can classical neural network architectures be systematically extended to handle inputs that are inherently quantum\cite{eisert2025mind, huang2025vast}?
Unlike classical networks, which operate on vectors or probability distributions, quantum data is natively represented by density operators.
In this regime, the features relevant for learning often correspond to the expectation values of non-commuting observables.
Bridging this gap requires architectures that not only respect the linearity of quantum evolution, but also navigate the high-dimensional geometry of quantum states to capture non-local correlations.

Several recent studies have sought to extend classical machine learning architectures to quantum settings. 
For example, hybrid quantum classical neural networks embed classical data into parameterized quantum circuits, leveraging variational ansätze for learning tasks \cite{schuld2020circuit, mari2020transfer}.
Other approaches adapt classical structural motifs\cite{beer2020training, li2026knowledge}, such as convolutional layers\cite{broughton2020tensorflow, cong2019quantum, ceschini2025hybrid}, recurrent architectures\cite{li2023quantum, takaki2021learning}, or residual connections\cite{liang2021hybrid, wen2024enhancing, kashif2024resqnets}, directly to quantum circuits.
These strategies preserve some of the inductive biases and optimization benefits of classical models while exploiting quantum feature spaces. 
Nonetheless, a gap remains, most existing methods either lack a formal correspondence with their classical counterparts or cannot systematically process fully quantum inputs, leaving open the question of how to construct architectures that are both consistent with classical networks and capable of capturing inherently quantum correlations.

In this work, we bridge these domains by introducing a Hybrid Quantum Residual Network (HQRN) architecture.
While standard quantum neural networks often suffer from optimization challenges like barren plateaus\cite{mcclean2018barren, larocca2025barren}, our HQRN provides a systematic extension of the classical residual learning motif.
When the inputs are restricted to the computational basis, the HQRN reduces to a classical network with residual connections\cite{he2016identity, he2016deep}. 
This equivalence implies that HQRN does not require independent training for classical tasks.
Instead, it can be initialized using optimized classical weights, effectively inheriting the convergence benefits of the classical regime.
Conversely, for general quantum inputs, the HQRN leverages non-local correlations and off-diagonal information that are fundamentally inaccessible to the classical counterpart.
Therefore, the HQRN framework not only preserves the favorable optimization landscapes of classical residual architectures, but also systematically extends them to the quantum domain.

We validate this framework through two distinct tasks: (i) the recognition of handwritten digits, serving as a classical benchmark to verify the model’s consistency with its classical counterpart, and (ii) the classification of bipartite entanglement in unknown mixed states, demonstrating that HQRN surpass its classical counterpart when handling quantum tasks.

\section{Pipeline}

In classical residual neural network (ResNet)\cite{he2016identity, he2016deep}, a residual block implements the mapping $z_{k} = z_{k-1}+\mathcal{F}(z_{k-1})$, where $z_{k-1}$ and $z_k$ denote the input and output of the $k\text{-th}$ residual block respectively, and $\mathcal{F}(\cdot)$ is a parameterized transformation.
To extend this structure to quantum regime, we consider inputs given by mixed states.
Let $\rho^{(k-1)}$ denotes the input of the $k\text{-th}$ block, where the superscript labels the layer index. 
In particular, $\rho^{(0)}$ corresponds to the initial input state.

\begin{figure}[t]
    \begin{center}
        \includegraphics[width=0.48\textwidth]{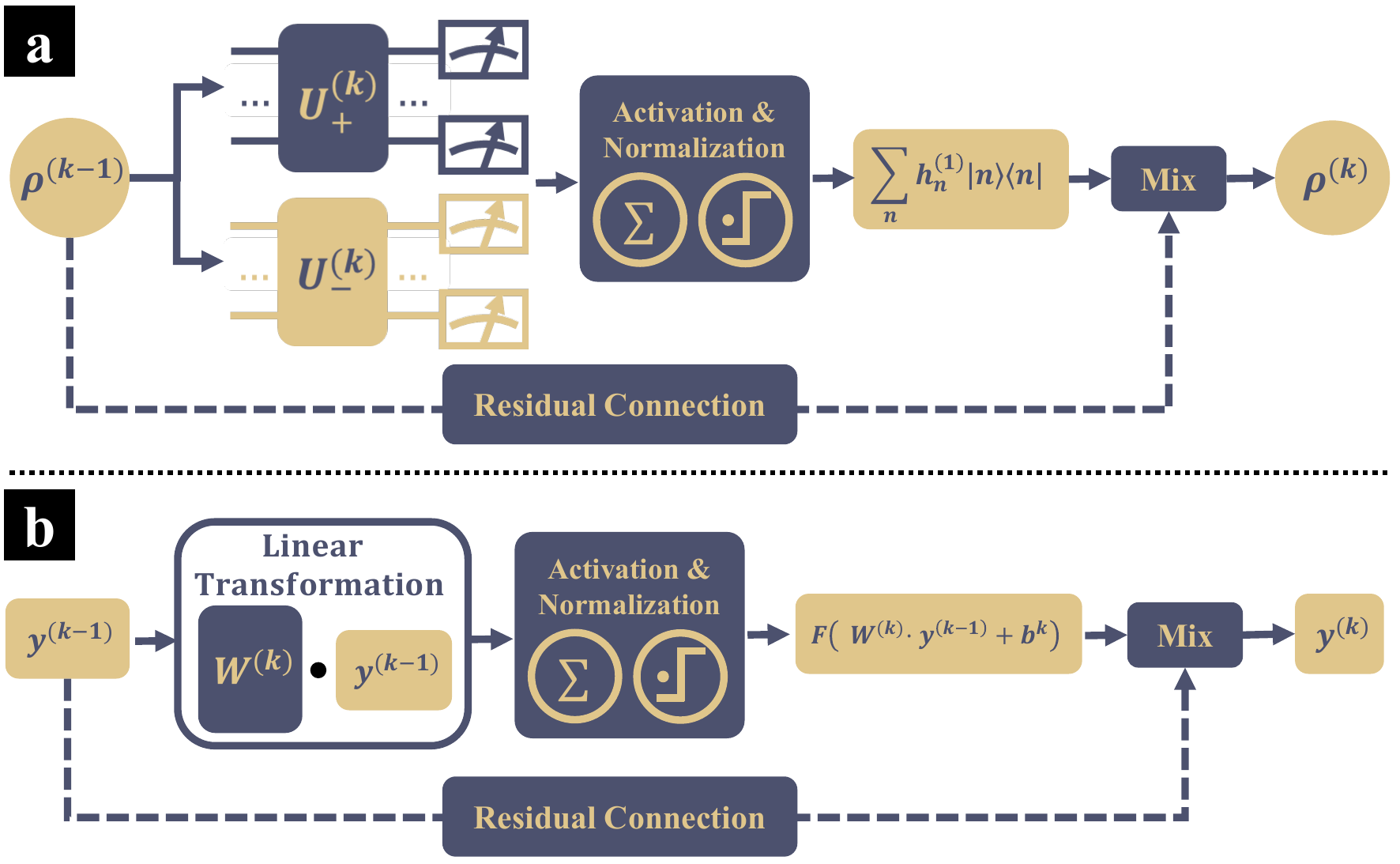}
    \end{center}
    \caption{
    {\bf Architecture of the $k$-th quantum residual block (QRB) in the proposed hybrid quantum classical residual network (HQRN). 
    }
    (a) Schematic of a single QRB.
    When the input is restricted to the computational basis (diagonal density matrices), the QRB reduces to a classical residual block (CRB) with weights $W_k$ and bias $b_k$. 
    In (b) we present this equivalent classical representation.
    }
    \label{fig_pipeline}
\end{figure}

Fig.(\ref{fig_pipeline}) illustrates the schematic of the $k\text{-th}$ quantum residual block (QRB) in a cascade.
First, we apply unitary operations $U^{(k)}_+$ and $U^{(k)}_-$ on the input state $\rho^{(k-1)}$.
The probability of obtaining result $n$ upon measuring the output is given by
\begin{equation}
    p^{(k)}_{\pm,n}=
    Tr\left(U^{(k)}_\pm\rho^{(k-1)}{U^{(k)}_\pm}^\dagger |n\rangle\langle n|\right)
\end{equation}
Next, we apply a non-negative activation function $f(\cdot)$ (e.g. ReLU, sigmoid\cite{pandey2019machine}) and normalize the results to derive the coefficients $h^{(k)}$.
For simplicity, we define a normalized non-linear mapping $F(\cdot)$ as
\begin{equation}
	F\left(z_n\right) \equiv \frac{f(z_n)}{\sum_lf(z_l)}
\end{equation}
and the coefficients $h^{(k)}$ can be given as
\begin{equation}
    h^{(k)}_n = F\left(\gamma^{(k)}(p^{(k)}_{+,n}-p^{(k)}_{-,n})+b^{(k)}_n\right)
    \label{eq_hk}
\end{equation}
where $\gamma^{(k)}$ and $b^{(k)}_n$ are the scaling factor and bias of the $k\text{-th}$ block respectively.
Eq.(\ref{eq_hk}) ensures that the resulting coefficients $\{h^{(k)}_n\}$ satisfy the unit-sum constraint for a density matrix.
With $h^{(k)}_n$, we then construct the intermediate state $\sum_nh^{(k)}_n|n\rangle\langle n|$.
The output of the $k\text{-th}$ block is obtained by mixing this state with the input $\rho^{(k-1)}$, yielding
\begin{equation}
    \rho^{(k)}=\alpha\rho^{(k-1)}+(1-\alpha)\sum_nh^{(k)}_n|n\rangle\langle n|
    \label{eq_rec}
\end{equation}
where $\alpha\in(0,1)$ is the residual weighting parameter.
By adjusting the parameter $\alpha$, this construction implements a tunable quantum residual connection that regulates the information flow from the input $\rho^{(k-1)}$ to the output $\rho^{(k)}$.
For clarity, the procedure for the $k\text{-th}$ block of the HQRN is summarized in Alg.(\ref{alg_res_block}).

\begin{algorithm}[H]
\caption{Pipeline of the $k\text{-th}$ QRB in HQRN}
\label{alg_res_block}
\begin{algorithmic}[1]
\Require Input state $\rho^{(k-1)}$, unitary operations $U^{(k)}_\pm$, scaling factor $\gamma^{(k)}$, bais $b^{(k)}$, non-negative activation function $f(\cdot)$, residual weighting parameter $\alpha\in(0, 1)$.
\State Apply $U^{(k)}_\pm$ on the input state $\rho^{(k-1)}$, then measure, collect results $p^{(k)}_{\pm,n}=Tr(U^{(k)}_\pm\rho^{(k-1)}{U^{(k)}_\pm}^\dagger|n\rangle\langle n|)$.
\State Apply activation function and normalize the results,  $h^{(k)}_n = F\left(\gamma^{(k)}(p^{(k)}_{+,n}-p^{(k)}_{-,n})+b^{(k)}_n\right)$, where $F(\cdot)$ is a normalized non-linear mapping, $F\left(z_n\right) \equiv {f(z_n)}/{\sum_lf(z_l)}$.
\State Construct new mixed state $\rho^{(k)}=\alpha\rho^{(k-1)}+(1-\alpha)\sum_nh^{(k)}_n|n\rangle\langle n|$.
\State \textbf{Output:} Mixed state $\rho^{(k)}$.
\end{algorithmic}
\end{algorithm}

Then we derive the output of $k$ cascading QRBs.
Assuming an arbitrary input state
\begin{equation}
    \rho^{(0)} = \sum_m h^{(0)}_m |\phi_m\rangle\langle \phi_m|
    \label{eqs_rho0}
\end{equation}
where $h^{(0)}_m \geq 0$ and $\sum_m h^{(0)}_m = 1$.
After $k$ successive QRBs, the output of the $k$-th block can be obtained by recursively applying Eq.(\ref{eq_rec}),
\begin{equation}
	\rho^{(k)} = 
	(1-\alpha)\sum_n
	\sum_{j=0}^{k-1}\alpha^jh_n^{(k-j)}
	|n\rangle\langle n| + \alpha^k\rho^{(0)}
    \label{eq_rhok}
\end{equation}
where $\sum_n h_n^{(j)} |n\rangle \langle n|$ represents the intermediate mixture generated by the $j$-th block.
The explicit form of coefficients $h_n^{(k)}$ in the $k$-th residual block is given by
\begin{widetext}
	\begin{equation}
		h_n^{(k)} = F\left(
		(1-\alpha)\sum_m
		\sum_{j=0}^{k-2}\alpha^jh_m^{(k-1-j)}
		W_{nm}^{(k)} + \alpha^{k-1}\sum_m\Omega_{nm}^{(k)}h_m^{(0)}
		+b_n^{(k)}
		\right)
        \label{eq_hnk}
	\end{equation}
\end{widetext}
where the weights are determined by $U_\pm$,
\begin{equation}
		\Omega^{(k)}_{nm} = 
        \gamma^{(k)}\left(
        \left|\langle m|U_+^{(k)}|\phi_n\rangle\right|^2
		- \left|\langle m|U_-^{(k)}|\phi_n\rangle\right|^2
        \right)
        \label{eq_omega}
\end{equation}
\begin{equation}
		W^{(k)}_{nm} = 
        \gamma^{(k)}\left(
        \left|\langle m|U_+^{(k)}|n\rangle\right|^2
		- \left|\langle m|U_-^{(k)}|n\rangle\right|^2
        \right)
        \label{eq_w}
\end{equation}
By optimizing the parameterized unitary operations $U_\pm$ within each residual block, we can tune the weights $W$ and $\Omega$ that govern the network's evolution.
In the limit of large depth, the recursive summation in  Eq.(\ref{eq_rhok}) and Eq.(\ref{eq_hnk}) allows the network to approximate any continuous mapping from the initial quantum state space to the diagonal probability simplex, analogous to the Universal Approximation Theorem for classical neural networks\cite{cybenko1989approximation, hornik1991approximation, lloyd1996universal}.
More details regarding the recursive expansion of $\rho^{(k)}$ are provided in the Supplemental Materials.

Interestingly, when inputs are restricted to the computational basis, HQRN is equivalent to a classical neural network with residual connections.
In this scenario, we denote the input to the $k$-th block (which is also the output of the $(k-1)$-th block) as $\tilde{\rho}^{(k-1)}=\sum_ny^{(k-1)}_n|n\rangle\langle n|$, where the tilde avoids confusion with the general case.
Within the $k$-th QRB, we still apply parameterized unitary operations $U^{(k)}_\pm$, perform measurements, and collect the resulting coefficients $\tilde{h}_n^{(k)}$.
As the input is already a diagonal mixed state, the weight matrices $\Omega$ and $W$ become identical.
Consequently, these coefficients simplifies to a standard linear transformation followed by a non-linear activation
\begin{equation}
    \tilde{h}_n^{(k)} = F\left(
    \sum_mW_{nm}^{(k)}y^{(k-1)}_m + b_n^{(k)}
    \right)
\end{equation}
The output is also a diagonal mixed state, obtained by mixing the intermediate state $\sum_n\tilde{h}_n^{(k)}|n\rangle\langle n|$ and the input
\begin{equation}
    \begin{split}
        \tilde{\rho}^{(k)}
        &=\sum_n
        \left(
        (1-\alpha)\tilde{h}_n^{(k)}
        +\alpha y^{(k-1)}_n
        \right)
        |n\rangle\langle n|
    \end{split}
\end{equation}
Therefore, the $k$-th QRB implements the recursive transformation
\begin{equation}
    y^{(k-1)}_n\longrightarrow 
    y^{(k)}_n = (1-\alpha)\tilde{h}_n^{(k)}
        +\alpha y^{(k-1)}_n
    \label{eq_update_y}
\end{equation}

By this mean, we establish a direct correspondence between the quantum state evolution and the discrete dynamics of classical residual connections.
It is important to note that while the QRB shares the fundamental recursive topology of classical residual networks, it does not strictly emulate the internal layer sequence of standard ResNet implementations.

\begin{figure*}[t]
    \begin{center}
        \includegraphics[width=0.85\textwidth]{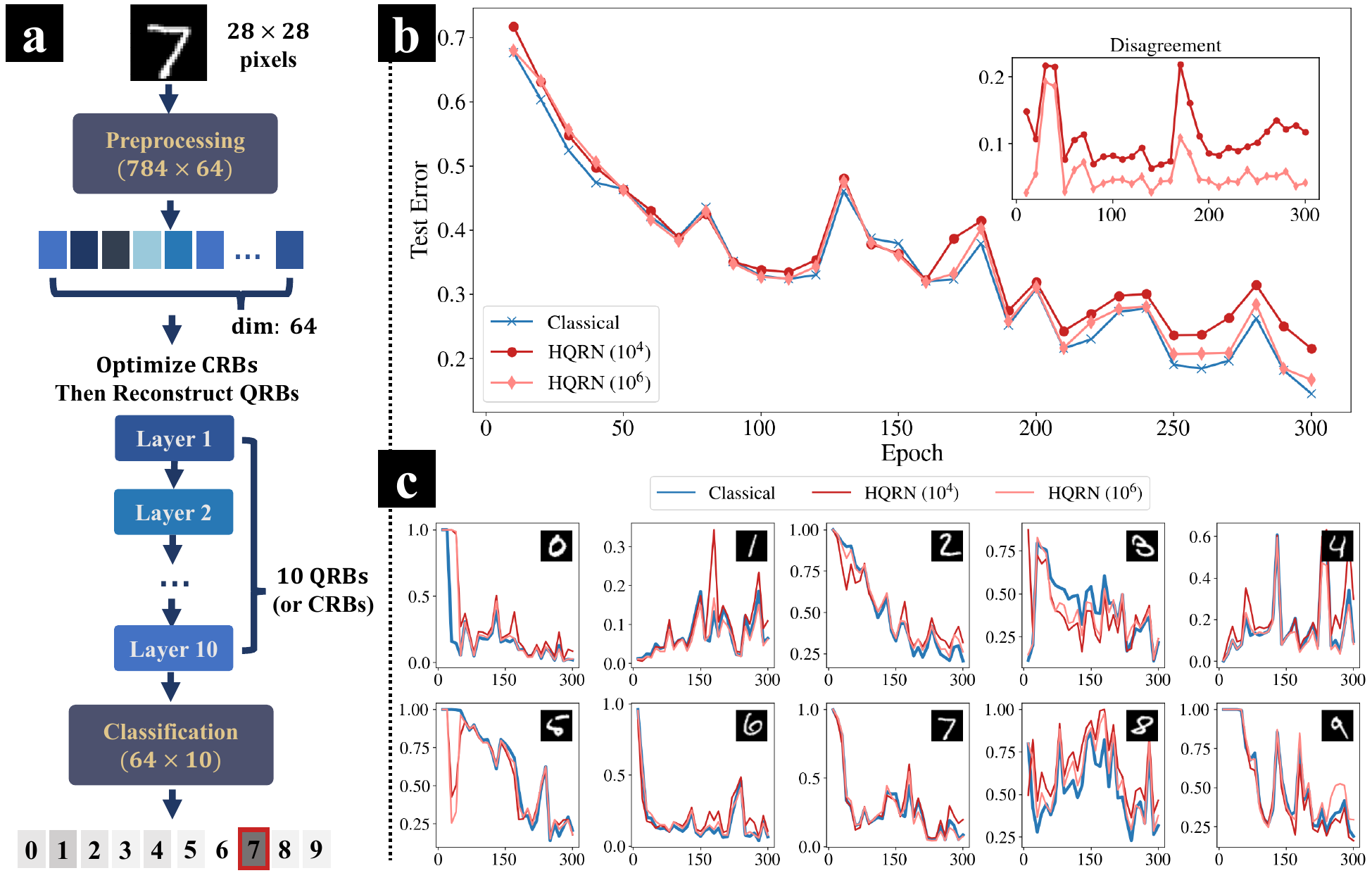}
    \end{center}
    \caption{
    {\bf Recognition of the MNIST handwritten digit dataset via HQRN.}
    (a) Schematic of the benchmark architecture. 
    (b) Comparison of the error rate across 300 training epochs for the classical baseline (blue) and the reconstructed HQRN with $N_s = 10^4$ (red) and $10^6$ (orange) measurement shots.
    The inset displays the disagreement rate between classical and quantum models. 
    (c) Class-wise error rate for digits 0--9, demonstrating the consistency of the HQRN across the entire feature space. 
    }
    \label{fig_digit}
\end{figure*}

\section{digit recognition}
\label{digit recognition}

We first evaluate the HQRN using the benchmark of handwritten digits recognition to verify the equivalence between the quantum and classical architectures when handling classical tasks.
The configuration of the benchmark is illustrated in Fig.(\ref{fig_digit}a), where we investigate supervised classification on the MNIST dataset\cite{lecun2002gradient, deng2012mnist}.
The input consists of single-channel, grayscale $28\times 28$ pixel images. 
After flattening the pixel intensities to the range $[0, 1]$, the data undergo a preprocessing stage where they are projected into a 64-dimensional probability vector space.
To ensure a valid mapping to the quantum state space, we enforce a non-negativity constraint and apply vector normalization.
The core of the architecture comprises a cascade of 10 successive residual blocks, as depicted in Fig.\ref{fig_pipeline}.
In this diagonal regime, the HQRN blocks recursively update the state according to Eq.(\ref{eq_update_y}).
By the end, the output of the terminal residual block is fed into a classification layer to obtain the predicted digit labels.

In the classical input regime, the functional equivalence between the HQRN and its classical analog allows for the direct mapping of classical weights onto quantum unitary operations.
We first optimize a classical residual network for 300 epochs.
The resulting convergence of the training error is illustrated by the blue curve in Fig.(\ref{fig_digit}b).
Upon the optimized weights $W$ and biases $b$ for each epoch, we then reconstruct the corresponding parameterized unitary operations $U_\pm$ through a combination of unitary dilation\cite{nagy2010harmonic} and Trotter decomposition\cite{lloyd2013quantum, ostmeyer2023optimised}. 
In this task, the $64\times64$ weight matrices are mapped to unitary operations on 7 qubits ($64=2^6$, we need one more qubit for dilation).
This protocol enables us to map the classical optimized weights directly onto the HQRN parameter space.

To evaluate the performance under realistic hardware constraints, we simulate the HQRN using finite measurement shots.
Denote $N_s$ as the number of measurement shots utilized to estimate the probabilities $p_+$ or $p_-$ of a single residual block.
For the first block, the reconstruction of the output state $\rho^{(1)}$ requires $4N_s$ copies of the initial input $\rho^{(0)}$. Specifically, $2N_s$ copies are consumed by the unitary rotations $U^{(1)}_\pm$ and subsequent measurement to compute the non-linear updates. The remaining $2N_s$ copies are stochastically mixed with the intermediate diagonal state to reconstruct $4N_s$ copies of the output $\rho^{(1)}$, maintaining the state ensemble size for the subsequent layer. This linear scaling of state copies per layer ensures that the HQRN maintains a tractable resource overhead even for deep architectures.

As shown in Fig.(\ref{fig_digit}b), the error rates for HQRN configurations with $N_s = 10^4$ (red) and $N_s = 10^6$ (orange) shots demonstrate that the quantum architecture successfully recovers the classical baseline as the sampling fluctuation is suppressed.
The inset of Fig.(\ref{fig_digit}b) quantifies the disagreement between the classical baseline and the reconstructed HQRN, defined by the frequency of divergent classification outcomes.
For $N_s = 10^4$ (red), the disagreement stabilizes at approximately $12\%$ after 250 epochs, whereas increasing the sampling budget to $N_s = 10^6$ (orange) reduces this discrepancy to approximately $3\%$.
Notably, we observe peaks in the disagreement during the $40\text{--}60$ and $160\text{--}180$ epoch intervals.
These peaks suggest that during optimization, the network resides in a sensitive region of the loss landscape, so that stochastic fluctuations in the quantum measurement process are significantly amplified.
Furthermore, Fig.(\ref{fig_digit}c) provides a class-wise breakdown of the error rates for each digit. 
The alignment between the HQRN results and the classical baseline confirms that the quantum architecture successfully recovers the functionality of the classical counterpart when operating on classical inputs.

\section{entanglement classifier}
\label{entanglement classifier}

\begin{figure*}[t]
    \begin{center}
        \includegraphics[width=0.95\textwidth]{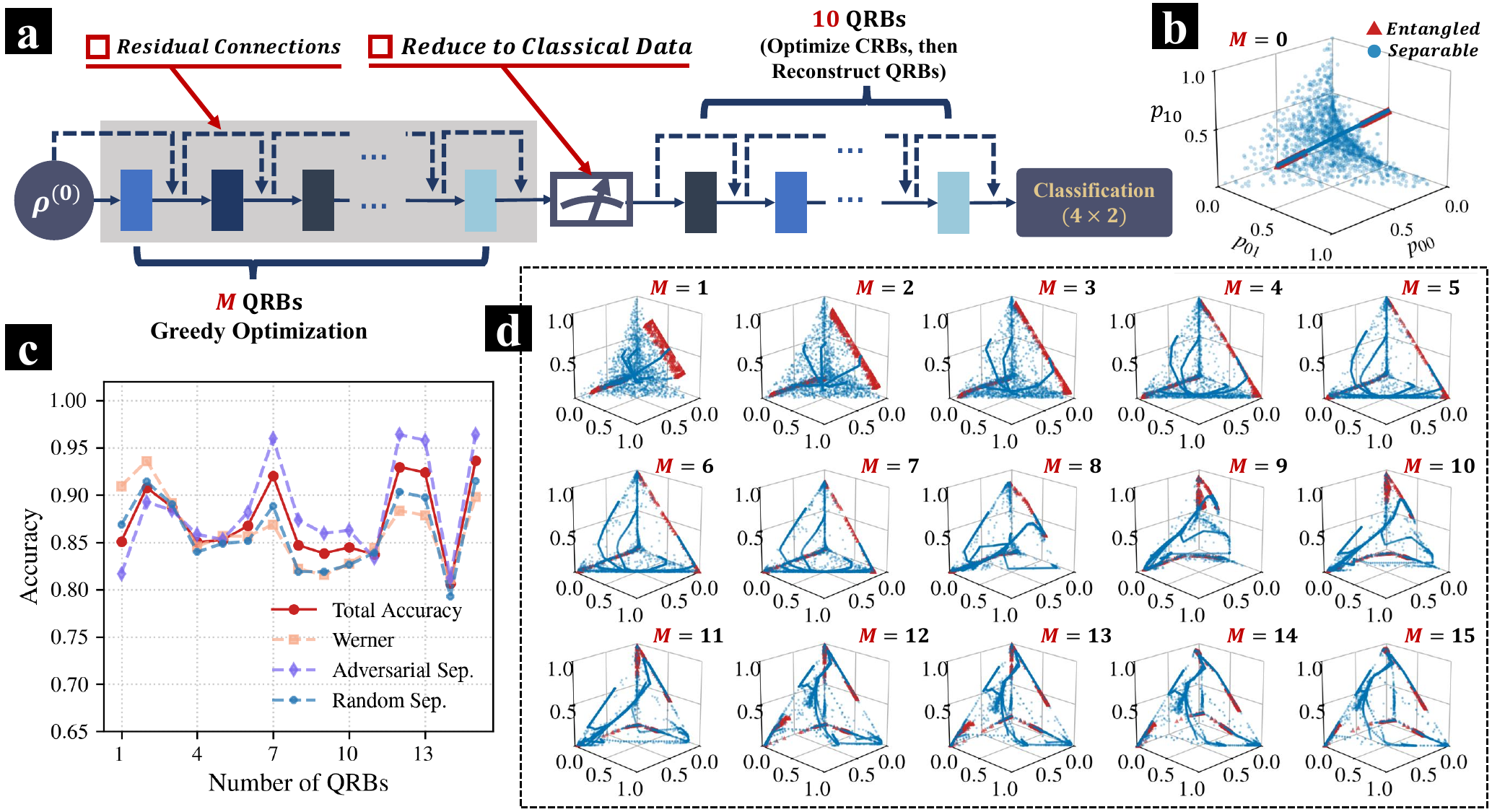}
    \end{center}
    \caption{
    {\bf Entanglement classification via HQRN.}
    (a) Schematic of the HQRN architecture. The input state $\rho^{(0)}$ evolves through a sequence of $M$ quantum residual blocks (QRBs). 
    Following Pauli-Z measurement, the resulting 4-dimensional probability vector is processed by a 10 QRBs for classification stage.
    (b) Direct measurement results of the test data ($M=0$) in the 3-simplex probability space $(p_{00}, p_{01}, p_{10})$.
    Entangled states (red triangles) and separable states (blue circles) exhibit spatial overlap. 
    (c) Classification accuracy against QRB depth $M$. 
    (d) Dynamical evolution of the intermediate measurement results from $M=1$ to $M=15$.
    The QRBs act as learned basis rotations, driving entangled and separable states along divergent trajectories to eliminate class overlap and maximize distinguishability.
    }
    \label{fig_ent}
\end{figure*}

To evaluate the capacity of HQRN to process quantum data, we investigate its performance as entanglement classifier.
We construct a comprehensive dataset $\mathcal{D} = \{(\rho_i, y_i)\}$, where $y_i = +1$ denotes entangled states and $y_i = -1$ denotes separable states.
The ensemble is designed to probe the boundary of the separable set through three distinct families, including Werner states, random separable mixed states, and adversarial separable states. 
The generalized two-qubit Werner states are defined by
\begin{equation}
    \rho_W(p_B) = p_B |\Phi_B\rangle\langle\Phi_B| + \frac{1-p_B}{4} \mathbb{I}
    \label{eq_werner}
\end{equation}
$|\Phi_B\rangle$ is one of the Bell states, $\{\frac{1}{\sqrt{2}}(|00\rangle + e^{i\varphi}|11\rangle, \frac{1}{\sqrt{2}}(|01\rangle + e^{i\varphi}|10\rangle)\}$ with a relative phase $\varphi \in [0, 2\pi)$.
Werner states\cite{werner1989quantum} are entangled (labeled as $+1$) for $p_B > 1/3$ and separable (labeled as $-1$) for $p_B \le 1/3$\cite{li2019entanglement, li2021quantum}.
We also generate random separable states $\rho_{\text{sep}} = \sum_k w_k (\rho_A^{(k)} \otimes \rho_B^{(k)})$, where $w_k\geq0$, and $\sum_kw_k=1$.
Additionally, we include adversarial separable states.
These "pseudo-entangled" states act as adversarial benchmarks by mimicking the marginal measurement statistics of Bell states.
They are constructed as $\rho_{ad} = p_{ad} \sigma_{ad} + (1-p_{ad}) \frac{\mathbb{I}}{4}$, where $\sigma_{ad}$ reproduces the same measurement results as a Bell state under certain measurements (e.g., Pauli-$X, Y,$ or $Z$), yet they are strictly separable.
$\rho_{ad}$ contains both mixed and pure states.
For instance, $\rho_{ad}$ includes pure states $\frac{1}{\sqrt{2}}(|00\rangle+|11\rangle)$, and mixed states $\frac{1}{2}(|0\rangle\langle 0|+|1\rangle\langle 1|)$.
Under Pauli-Z measurement, these states lead to same measurement results as Bell state $\frac{1}{\sqrt{2}}(|00\rangle + e^{i\varphi}|11\rangle$).

We implement a hybrid network as illustrated in Fig.(\ref{fig_ent}a).
The input state $\rho^{(0)}$ is processed by a cascade of $M$ QRBs, where dashed lines indicate the skip connections.
The $M$-th output $\rho^{(M)}$ is subjected to Pauli-Z measurement, yielding a 4D probability vector $\mathbf{p} = (p_{00}, p_{01}, p_{10}, p_{11})$., corresponding to the probability of obtaining results $|00\rangle$,$|01\rangle$,$|10\rangle$,$|11\rangle$.
This vector is subsequently processed by 10 QRBs.
As $\mathbf{p}$ can be mapped as a diagonal density matrix, here we instead train the equivalent CRBs, then reconstruct the 10 QRBs via unitary dilation and Trotter decomposition.
By the end, the last layer classifies whether the state is entangled or separable.

To circumvent the challenges of training deep quantum circuits, we employ a greedy layer-wise optimization strategy to the first $M$ QRBs.
The first $M$ QRBs are optimized to maximize the spatial distinguishability of the two classes within the simplex, while the succeeding 10 blocks are trained to minimize the final classification cross-entropy.

In numerical simulation, we use 1300 states as training data (including 350 Werner states, 300 random separable states, and 650 adversarial separable states).
And we use 5000 states as test data (including 1000 Werner states, 1500 random separable states, and 2500 adversarial separable states).
In this hybrid network, we apply ReLU activation function, and Pauli-Z measurements.
As we are studying bipartite system, each of the first $M$ QRBs contains 2 qubits.
After we measure the output of the $M$-th QRB, and encode the outputs as a 4-d vector on simplex, which can be mapped as a diagonal density matrix, we use 3 qubits in each QRB succeeding, as we need to implement unitary dilation to reconstruct the $4\times 4$ weights.

In Fig.(\ref{fig_ent}b,d), we present the intermediate measurement results after $M$ cascading QRBs.
As shown in Fig.(\ref{fig_ent}b), a direct Pauli-$Z$ measurement on the input states results in  overlap between the entangled (red triangles) and separable (blue circles) distributions. 
Due to the existence of adversarial states, adversarial states occupy the same region in the probability simplex.
Thus, classical networks can no more achieve perfect classification from the single Pauli-Z measurements results.
In Fig.(\ref{fig_ent}d), we present the evolution of $p_{00},p_{01},p_{10}$ within the 3-simplex probability space.
At low depth ($M=1$), it is still difficult to classify the entangled (red triangles) and separable (blue circles) ensembles, particularly around the $p_{01}=p_{10}=0$ axis, where there are both adversarial states and Werner states. 
As $M$ increases, the QRBs are optimized to drive the inputs along divergent trajectories, improving the spatial distinguishability of entangled and separable states.

In Fig.(\ref{fig_ent}c), we present the classification accuracy with various number of QRBs.
We can observe peaks around $M=2,7,12,13,15$, where the adversarial states are transformed away from the entangled ones, leading to a accurate classification of these adversarial states.
Notably, the total accuracy (red solid line) does not increase strictly monotonically. 
Because the first $M$ layers are optimized via a greedy strategy, the network tends to settle into local optima.
At intermediate depths (for example, $M = 14$), the entangled and separable manifolds may undergo temporary geometric compression. While the QRBs act as filters to segregate the classes, the states can become trapped in restricted regions of the probability simplex where the inter-class margins are insufficient for clear discrimination.
In these regimes, the measurement distributions may transiently overlap, obscuring the underlying quantum correlations. 
As the depth increases further, the succeeding residual blocks provide the necessary non-linear resolution to unfold these manifolds and restore classification fidelity.

This capacity to resolve overlapping manifolds, where its classical counterpart fails, is fundamentally rooted in the HQRN’s ability to extract a richer set of features from the input mixed states.
As established in Eqs. (\ref{eq_rhok}) and (\ref{eq_hnk}), the accumulation of $M$ successive QRBs creates a recursive transformation where each intermediate coefficient $h_n^{(k)}$ acts as a collector of results from a diverse branch of quantum measurements.
The weight matrix $\Omega^{(k)}_{nm}$, defined in Eq. (\ref{eq_omega}), encapsulates the projection probabilities of the initial state after the unitary operations $U^{(k)}_\pm$.
In the classical regime, $\rho^{(0)}$ is diagonal in the computational basis ($|\phi_m\rangle = |m\rangle$), and $\Omega^{(k)}_{nm}$ reduces to the classical weight matrix $W^{(k)}_{nm}$.
Thus, the information gain is bounded by the initial measurements. 
However, for general mixed states, such as the Werner states, the off-diagonal elements of $\rho^{(0)}$ are essential.
In HQRN, each QRB performs a learned "basis rotation" that rotates these off-diagonal correlations into the diagonal subspace of the succeeding layer.
As the depth $M$ increases, the HQRN aggregates these diverse projection results, effectively "unfolding" information that is invisible to cascading CRBs limited to insufficient measurements.
This recursive extraction of off-diagonal information allows the HQRN to resolve the manifolds of entangled and adversarial states, surpassing the limits of its classical counterpart.

\section{conclusions}
\label{conclusions}

In conclusion, our work establishes the Hybrid Quantum-Residual Network (HQRN), as a versatile architecture that provides exact functional equivalence between classical residual learning and quantum state evolution.
We established that for inputs characterized by diagonal density operators, the HQRN exhibits an exact functional equivalent to classical networks with residual connections.
This symmetry enables quantum models to directly inherit classical optimization benefits, effectively bypassing the challenges associated with deep quantum circuit optimization.
Moreover, we show that the HQRN is more than a quantum emulator.
Instead, it is a conceptual extension that thrives in the quantum regime.
In the entanglement classification task, HQRN achieves high classification accuracy even for adversarial separable states that mimic the marginal measurement statistics of entangled pairs.
This framework offers a scalable, resource-efficient template for deep quantum networks on near-term quantum hardware and beyond.

{\it Acknowledgments} 
J.L gratefully acknowledges funding by National Natural Science Foundation of China (NSFC) under Grant No.12305012.

\bibliography{ref}

\clearpage 
\onecolumngrid 
\begin{center}
    \textbf{\large Supplementary Materials}
\end{center}

\beginsupplement 

\section{The Output of Successive Residual Blocks}

In the Hybrid quantum residual network (HQRN) architecture, the output of the $k\text{-th}$ quantum residual block (QRB) is obtained by mixing intermediate state $\sum_nh^{(k)}_n|n\rangle\langle n|$ with the input $\rho^{(k-1)}$,yielding
\begin{equation}
    \rho^{(k)}=\alpha\rho^{(k-1)}+(1-\alpha)\sum_nh^{(k)}_n|n\rangle\langle n|
    \label{eqs_resc}
\end{equation}
where $\alpha\in(0,1)$ is the residual weighting parameter.
In this section, we derive the exact expression for the output state after $k$ residual blocks, with arbitrary initial input
\begin{equation}
    \rho^{(0)} = \sum_m h^{(0)}_m |\phi_m\rangle\langle \phi_m|
    \label{eqs_rho0}
\end{equation}
where $h^{(0)}_n\geq0$, and $\sum_nh^{(0)}_n=1$. 

In the first block, unitary operations $U_\pm^{(1)}$ are applied to the input $\rho^{(0)}$.
The probability of obtaining result $n$ upon measuring the output is given by
\begin{equation}
    p^{(1)}_{\pm,n}=
    Tr\left(U^{(1)}_\pm\rho^{(0)}{U^{(1)}_\pm}^\dagger |n\rangle\langle n|\right)
\end{equation}
Substituting Eq.(\ref{eqs_rho0}), we have
\begin{equation}
    p^{(1)}_{\pm,n}=\sum_m h^{(0)}_m 
    \left|
    \langle n|U^{(1)}_\pm|\phi_m\rangle
    \right|^2
\end{equation}
Next, we apply a non-negative activation function $f(\cdot)$ and normalize the results to derive the hidden state
\begin{equation}
    h^{(1)}_n=\frac{
    f\left(\gamma^{(1)}(p^{(1)}_{+,n}-p^{(1)}_{-,n})+b^{(1)}_n\right)}
    {\sum_jf\left(\gamma^{(1)}(p^{(1)}_{+,j}-p^{(1)}_{-,j})+b^{(1)}_j\right)}
    \label{eqs_h1}
\end{equation}
where $\gamma^{(1)}$, $b^{(1)}$ are the scaling factor and bias of the first block, respectively.
For simplicity, we define a normalized non-linear mapping
\begin{equation}
	F\left(z_n\right) = \frac{f(z_n)}{\sum_lf(z_l)}
\end{equation}
where $f(\cdot)$ is the non-negative activation function, and the denominator ensures the normalization of the output probabilities.
Recalling the input state in Eq.(\ref{eqs_rho0}), Eq.(\ref{eqs_h1}) can be rewritten as
\begin{equation}
    h^{(1)}_n
    =
    F\left(
	\sum_m \Omega^{(1)}_{nm} h^{(0)}_m + b_n^{(1)}
	\right)
    \label{eqs_h1_sim}
\end{equation}
where
\begin{equation}
	\Omega^{(1)}_{nm} = \gamma^{(1)}\left(
    \left|\langle n|U_+^{(1)}|\phi_m\rangle\right|^2
	- \left|\langle n|U_-^{(1)}|\phi_m\rangle\right|^2
    \right)
\end{equation}
Substituting Eq.(\ref{eqs_h1}) to Eq.(\ref{eqs_resc}), we obtain the output of the first block
\begin{equation}
    \begin{split}
        \rho^{(1)} 
        &= (1-\alpha)h^{(1)}_n|n\rangle\langle n| + \alpha\rho^{(0)}
        \\
        &= (1-\alpha) F\left(
	      \sum_n \Omega^{(1)}_{nm} h^{(0)}_n + b_n^{(1)}
	    \right)
	    |n\rangle\langle n| + \alpha\rho^{(0)}
	\label{eq_rho1}
    \end{split}
\end{equation}

In the second block, $\rho^{(1)}$ is input.
Similarly, unitary operations $U_\pm^{(2)}$ are applied to $\rho^{(1)}$, and the probability of obtaining result $n$ upon measuring the output is given by
\begin{equation}
    \begin{split}
        p^{(2)}_{\pm,n} 
        =&
        Tr\left(U^{(1)}_\pm\rho^{(1)}{U^{(2)}_\pm}^\dagger |n\rangle\langle n|\right)
        \\=&
        (1-\alpha)\sum_m h^{(1)}_m \left|\langle n|U^{(2)}_\pm|m\rangle\right|^2
        \\
        &+\alpha\sum_m h^{(0)}_m 
        \left|
        \langle n|U^{(1)}_\pm|\phi_m\rangle
        \right|^2
    \end{split}
    \label{eqs_p2}
\end{equation}
In Eq.(\ref{eqs_p2}), the first term arises from the intermediate state $\sum_n h_n^{(1)}|n\rangle\langle n|$, while the second term represents the contribution from the preceding state $\rho^{(1)}$.

Following the same procedure, the hidden state $h^{(2)}_n$ is determined by applying the normalized non-linear mapping $F(\cdot)$ to the measurement outcomes of the second block, 
	 
		\begin{equation}
			h_n^{(2)} = F\left(
			\alpha \sum_m \Omega^{(2)}_{nm} h^{(0)}_m 
			+ (1-\alpha) \sum_m W^{(2)}_{nm} h^{(1)}_m + b_n^{(2)}
			\right)
		\end{equation}
	 
where for simplicity, we denote 
\begin{equation}
	\Omega^{(2)}_{nm} = \gamma^{(2)}\left(
    \left|\langle n|U_+^{(2)}|\phi_m\rangle\right|^2
	- \left|\langle n|U_-^{(2)}|\phi_m\rangle\right|^2
    \right)
\end{equation}
\begin{equation}
	W^{(2)}_{nm} = \gamma^{(2)}\left(
    \left|\langle n|U_+^{(2)}|m\rangle\right|^2
	- \left|\langle n|U_-^{(2)}|m\rangle\right|^2
    \right)
\end{equation}
Following the recursive construction in Eq.~(\ref{eqs_resc}), the output of the second block is obtained by mixing the intermediate state $\sum_nh_n^{(2)}|n\rangle\langle n|$ with the previous output $\rho^{(1)}$, yielding
\begin{equation}
    \begin{split}
        \rho^{(2)} 
        &= (1-\alpha)\sum_n h^{(2)}_n|n\rangle\langle n| + \alpha \rho^{(1)}
        \\
        &= (1-\alpha)\sum_n 
		\left( h^{(2)}_n + \alpha  h^{(1)}_n \right)
		|n\rangle\langle n| + \alpha^2 \rho^{(0)}
    \end{split}
\end{equation}

By applying this procedure recursively, we obtain the output of the $k$-th block as
 
	\begin{equation}
			\rho^{(k)} = 
			(1-\alpha)\sum_n
			\left(
			h_n^{(k)} + \alpha h_n^{(k-1)} + \alpha^2 h_n^{(k-2)}
			+\cdots+ \alpha^{k-1} h_n^{(1)}
			\right)|n\rangle\langle n| + \alpha^k\rho^{(0)}
	\end{equation}
 
	where
	 
		\begin{equation}
			h_n^{(k)} = F\left(
			(1-\alpha)\sum_m
			\left(
			h_m^{(k -1)} + \alpha h_m^{(k-2)} + \alpha^2 h_m^{(k-3)}
			+\cdots+ \alpha^{k-2} h_m^{(1)}
			\right)W_{nm}^{(k)} + \alpha^{k-1}\sum_m\Omega_{nm}^{(k)}h_m^{(0)}
			+b_n^{(k)}
			\right)
		\end{equation}
	 
where the weights are given as
	\begin{equation}
		\Omega^{(k)}_{nm} = \left|\langle m|U_+^{(k)}|\phi_n\rangle\right|^2
		- \left|\langle m|U_-^{(k)}|\phi_n\rangle\right|^2
	\end{equation}
	\begin{equation}
		W^{(k)}_{nm} = \left|\langle m|U_+^{(k)}|n\rangle\right|^2
		- \left|\langle m|U_-^{(k)}|n\rangle\right|^2
	\end{equation}
For simplicity, $\rho^{(k)}$ can be rewritten as
	\begin{equation}
		\rho^{(k)} = 
		(1-\alpha)\sum_n
		\sum_{j=0}^{k-1}\alpha^jh_n^{(k-j)}
		|n\rangle\langle n| + \alpha^k\rho^{(0)}
	\end{equation}
and coefficients $h_n^{(k)}$ can be rewritten as
	 
		\begin{equation}
			h_n^{(k)} = F\left(
			(1-\alpha)\sum_m
			\sum_{j=0}^{k-2}\alpha^jh_m^{(k-1-j)}
			W_{nm}^{(k)} + \alpha^{k-1}\sum_m\Omega_{nm}^{(k)}h_m^{(0)}
			+b_n^{(k)}
			\right)
		\end{equation}
        \label{eqs_h_exact}

Specifically, for non-negative and normalized classical inputs ${\bf x}$ satisfying $x_j \geq 0$ and $\sum_j x_j = 1$, we encode the data as a diagonal mixed state
\begin{equation}
    \tilde{\rho}^{(0)} = \sum_{n=1}^{2^N}x_n|n\rangle\langle n|
\end{equation}
where tilde notation indicates classical inputs.
The input space is constrained to the $(2^N-1)$-dimensional probability simplex, ensuring that the classical vector ${\bf x}$ maps directly to the eigenvalues of a valid, unit-trace density matrix $\tilde{\rho}^{(0)}$.
In this case, the weight matrices $\Omega$ and $W$ become identical. 
Consequently, the output of each residual block becomes diagonal mixed state that shapes as
\begin{equation}
    \tilde{\rho}^{(k)} 
    = 
    \sum_n
    \left(
        (1-\alpha)\sum_{j=0}^{k-1}\alpha^j\tilde{h}_n^{(k-j)}
        +
        \alpha^k x_n
    \right)
		|n\rangle\langle n|
\end{equation}
and the coefficients $\tilde{h}_n^{(k)}$ for the $k$-th residual block reduce to a standard linear transformation followed by a non-linear activation
\begin{equation}
    \tilde{h}_n^{(k)} = F\left(
    \sum_mW_{nm}^{(k)}y^{(k-1)}_m + b_n^{(k)}
    \right)
    \label{eqs_h_classical}
\end{equation}
and the exact form can be derived from Eq.(\ref{eqs_h_exact}),
 
    \begin{equation}
			\tilde{h}_n^{(k)} = F\left(
			\sum_m
            W_{nm}^{(k)}
			\left(
            \sum_{j=0}^{k-2}
            (1-\alpha)\alpha^j\tilde{h}_m^{(k-1-j)}
			 + \alpha^{k-1}x_m
             \right)
			+b_n^{(k)}
			\right)
            \label{eqs_h_exact_classical}
\end{equation}

\section{Implementation and Optimization Details for Digit Recognition}

\begin{figure}[t]
    \begin{center}
        \includegraphics[width=0.48\textwidth]{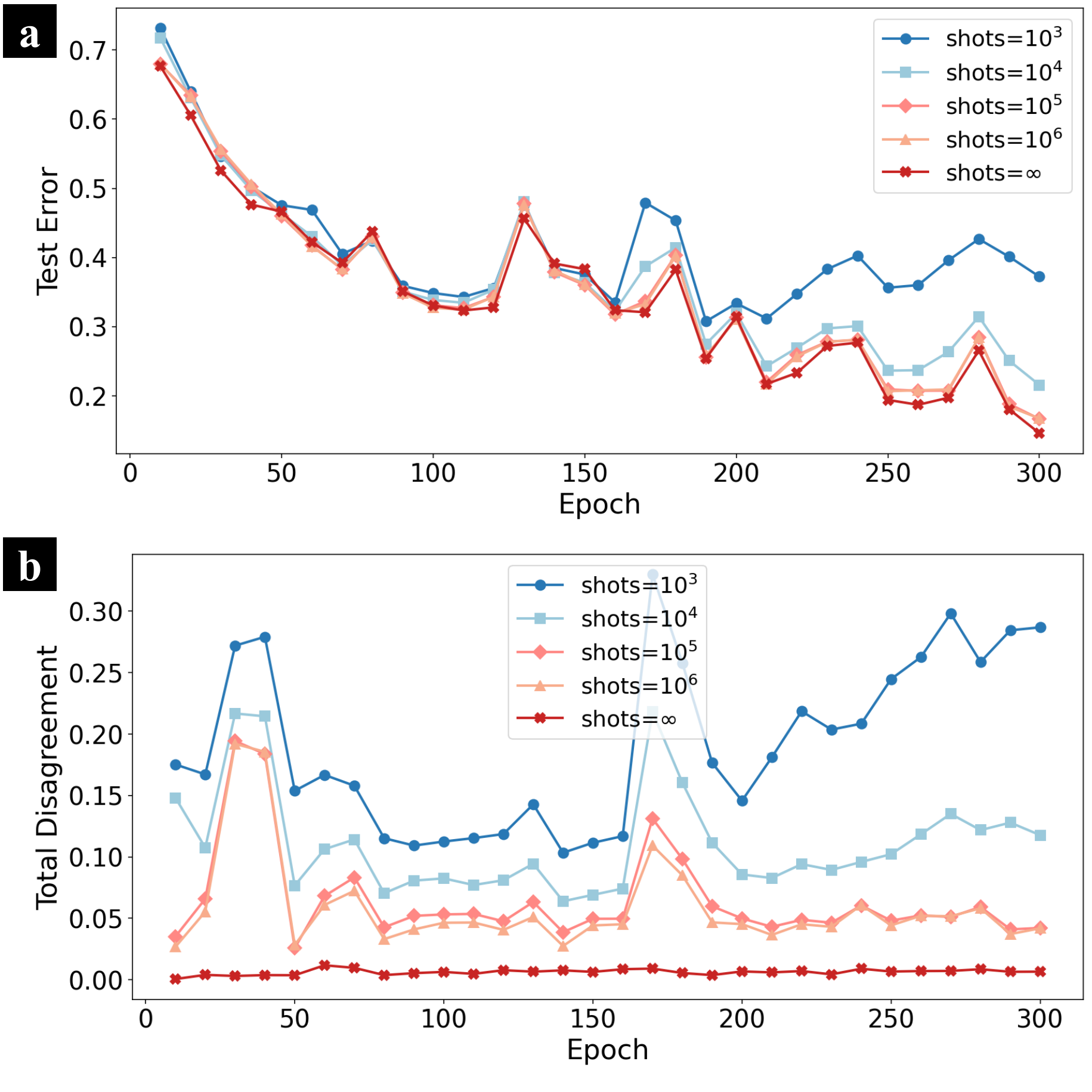}
    \end{center}
    \caption{
    {\bf Test error and total model disagreement across various shots.} 
    (a) Test error evolution over 300 epochs for the HQRN reconstructed model varying shots $N_s \in \{10^3, 10^4, 10^5, 10^6, \infty\}$.
    Higher shot counts demonstrate monotonic convergence toward the classical result.
    (b) Total disagreement rate (defined as the frequency of divergent classification labels between the classical baseline and the HQRN).
    }
    \label{figs_full_shots}
\end{figure}

\begin{figure*}[t]
    \begin{center}
        \includegraphics[width=0.98\textwidth]{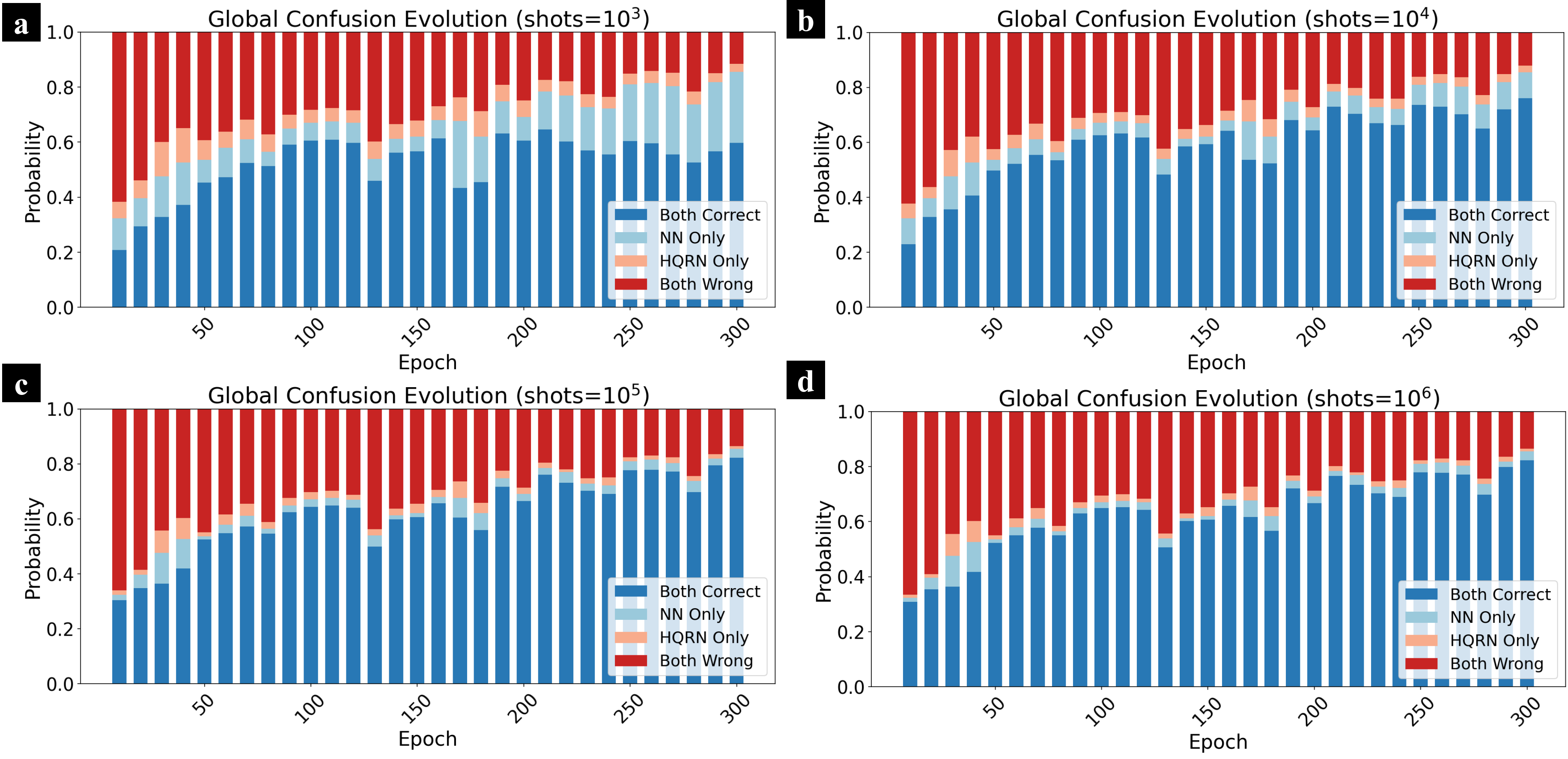}
    \end{center}
    \caption{
    {\bf Global Confusion Evolution across different shot}
    The four panels (a–d) illustrate the classification dynamics for $N_s = 10^3, 10^4, 10^5,$ and $10^6$, respectively. The stacked probability plots track four distinct outcomes: mutual correct classification (Both Correct, light blue), mutual incorrect classification (Both Wrong, purple), and divergent outcomes where only one model is correct (NN Only, dark blue; HQRN Only, orange). 
    }
    \label{figs_confusion}
\end{figure*}

In this section, we provide the full derivation and implementation details for the recognition of handwriting digits.

In digit recognition, each grayscale input image $x \in \mathbb{R}^{1\times 28 \times 28}$ is initially mapped to a probability simplex.
To achieve this, we flatten the input into a vector of dimension $D=784$ and apply the transformation
\begin{equation}
\tilde{x}_i = \frac{\max(x_i,0)}{\sum_j \max(x_j,0) + \varepsilon},
\end{equation}
where $\varepsilon > 0$ is a small constant ensuring numerical stability and strict adherence to the simplex constraints of nonnegativity and unit normalization.
In the numerical simulation, we set $\varepsilon = 10^{-20}$.

The learning architecture is constructed as depicted in Fig.(2a).
The model begins with an input projection $\mathbb{R}^{784} \to \mathbb{R}^{64}$, followed by a cascade of $10$ residual blocks as shown in Fig.(1b).
Each block implements the recursive transformation
\begin{equation}
    y^{(k-1)}_n\longrightarrow 
    y^{(k)}_n = (1-\alpha)\tilde{h}_n^{(k)}
        +\alpha y^{(k-1)}_n
    \label{eqs_update_y}
\end{equation}
where $y^{(k)}$ is the output of the $k{\text{-th}}$ block, the coefficients $\tilde{h}_n^{(k)}$ are given in Eq.(\ref{eqs_h_exact}) and Eq.(\ref{eqs_h_exact_classical}), and $\alpha\in(0,1)$ is the residual weighting parameter. 

In numerical simulation, Here we set $\alpha=0.5$, and use the ReLU activation function as $f(\cdot)$.
Thus, the $k{\text{-th}}$ block implements transformation
 
		\begin{equation}
y^{(k-1)}_n\longrightarrow 
y^{(k)}_n = \frac{1}{2}
    F\left(
    \sum_mW_{nm}^{(k)}y^{(k-1)}_m + b_n^{(k)}
    \right)
        +\frac{1}{2} y^{(k-1)}_n
\end{equation}
        \label{eqs_trans}
	 
where $F(\cdot)$ acts as
\begin{equation}
    \begin{split}
        &F\left(
    \sum_mW_{nm}^{(k)}y^{(k-1)}_m + b_n^{(k)}
    \right)
    \\
    =& 
    \frac{ReLU\left(
    \sum_mW_{nm}^{(k)}y^{(k-1)}_m + b_n^{(k)}
    \right)}{\sum_j{ReLU\left(
    \sum_mW_{jm}^{(k)}y^{(k-1)}_m + b_j^{(k)}
    \right)}}
    \end{split}
\end{equation}
Since the activation function $F(\cdot)$ enforces a normalization constraint on non-negative outputs, and given that the preceding state $y^{(k-1)}$ is by definition a normalized, non-negative vector, the recursive transformation ensures that the output $y^{(k)}$ remains strictly within the probability simplex. This property preserves the physical validity of the reconstructed density matrix at every depth.
We train the network on $10\%$ of the MNIST training set using the RMSProp optimizer with a learning rate of $3\times 10^{-3}$ and weight decay of $10^{-4}$.
The optimization objective is the minimization of the standard cross-entropy loss.
\begin{equation}
\mathcal{L} = -\sum_{j} \sum_{c} q_{j,c} \ln(\hat{y}_{j,c})
\end{equation}
where $q_{j,c}$ is the binary indicator (0 or 1) if class label $c$ is the correct classification for observation $i$, and $\hat{y}_{i,c}$ represents the predicted probability (the output of the final classification layer).

Following the optimization of the classical residual blocks, we reconstruct the corresponding HQRN by mapping the trained weights $W^{(k)}$ onto the parameterized unitary operators $U_{\pm}^{(k)}$.
This is achieved through a unitary dilation protocol, along with Trotter decomposition.
Specifically, for each block $k$, the weights $W_{nm}^{(k)}$ define the transition amplitudes between the input state and the measurement results. 
We construct $U_{\pm}^{(k)}$ such that the action of the quantum circuit, followed by computational basis measurements, recovers the linear transformation $\sum_m W_{nm}^{(k)} y_m^{(k-1)}$. 
Each weight matrix is decomposed into a positive and negative pair:
\begin{equation}
W = \lambda (W_P - W_N),
\end{equation}
where the positive and negative components are defined as $W_P = \frac{1}{2}(|W| + W)$ and $W_N = \frac{1}{2}(|W| - W)$.
By introducing a scaling factor $\lambda \ge 0$, we ensure that the spectral norms of $W_P$ and $W_N$ are bounded by unit.
Next, we parameterize them as $W_P = c |M_+|^2$ and $W_N = c |M_-|^2$, where $\|M_\pm\|_2 \leq 1$.
These contractive operators $M$ are then embedded into a higher-dimensional unitary space via unitary dilation (Speciafically, here we apply the standard Halmos dilation):
\begin{equation}
U(M) =
\begin{pmatrix}
M & \sqrt{I - M M^\dagger} \\
\sqrt{I - M^\dagger M} & -M^\dagger
\end{pmatrix}.
\end{equation}

To represent $U(M)$ on a quantum processor, we derive the corresponding representation by Trotter--Suzuki decomposition.
Assuming that the evolution operator $U(M)$ is decomposed into $r$ steps of a $p$-th order formula $S_p(t/r)$. 
Higher-order approximations are generated recursively:
\begin{equation}
S_{2k}(t) = S_{2k-2}(p_k t)\, S_{2k-2}((1-2p_k)t)\, S_{2k-2}(p_k t),
\end{equation}
with the coefficient $p_k = (4 - 4^{1/(2k-1)})^{-1}$. 
For each layer and training checkpoint, we execute a complete reconstruction pipeline consisting of weight extraction, decomposition into $M_\pm$ contractions, unitary embedding, and Trotterization.
The final reconstructed map is given by:
\begin{equation}
W_{\mathrm{rec}} = \lambda c \left(M_+^2 - M_-^2\right).
\end{equation}
where $\lambda c$ forms the $\gamma$ in this block.

To further investigate the performance of HQRN, we present the test error and total model disagreement across various shots in Fig.(\ref{figs_full_shots}).
As illustrated in Fig.(\ref{figs_full_shots}a), the test error for various shot counts $N_s$ follows the general trajectory of the idela limit ($N_s = \infty$), with convergence improving as the shots increases.
While the $N_s = 10^3$ case exhibits significant fluctuations, particularly in later epochs, the $N_s = 10^6$ trajectory is close to the ideal result, confirming the validity of reconstructed weights via unitary dilation and Trotter decompositions.
The total disagreement presented in Fig.(\ref{figs_full_shots}b) offers deeper insight into the network's stability.
Here disagreement describes the frequency of divergent classification outcomes between the classical model and its quantum reconstruction.
Even at high shot counts, we observe transient instability peaks, near epochs 40 and 175, where the disagreement spikes regardless of the sampling budget. 
These peaks likely correspond to critical points in the optimization landscape where the classical weights undergo rapid updates, rendering the resulting quantum mapping highly sensitive to stochastic fluctuations
. As the network stabilizes in the latter half of the training, the disagreement for $N_s = 10^6$ falls below $5\%$, indicating a high-fidelity recovery of the classical decision boundaries.

Additionally, we analyze the global confusion evolution across four sampling regimes in Fig. \ref{figs_confusion}.
These stacked probability plots reveal the composition of the classification outcomes throughout the training process.
At the lowest sampling budget $N_s = 10^3$, as shown in Fig.(\ref{figs_confusion}d), the "Both Correct" region is significantly constrained by divergent outcomes ("NN Only" and "HQRN Only"), which account for a substantial portion of the probability space.
This suggests that at low $N_s$, the fluctuations due to limit shots is sufficient to push the state across the decision boundary, leading to inconsistent labeling.
As $N_s$ increases to $10^6$, as shown in Fig.(\ref{figs_confusion}d), the divergent regions contract significantly, and the "Both Correct" area dominates, mirroring the classical convergence.
Crucially, the "Both Wrong" region (purple) persists and tracks the classical error rate closely across all shot regimes.
The reduction of the "HQRN Only" and "NN Only" bands toward the terminal epochs demonstrates that with sufficient measurement statistics, reconstructed HQRN is capable to achieve functional parity with the classical benchmark.

\section{Implementation and Optimization Details for Digit Recognition}

In this section, we provide the explicit construction of the Quantum Residual Blocks (QRBs) and the optimization protocol utilized to train the Hybrid Quantum-Residual Network (HQRN).

To evaluate the classification robustness of the HQRN, we construct a comprehensive dataset $\mathcal{D} = \{(\rho_i, y_i)\}$, where $y_i = +1$ denotes entangled states and $y_i = -1$ denotes separable states.

\begin{figure*}[t]
    \begin{center}
        \includegraphics[width=0.98\textwidth]{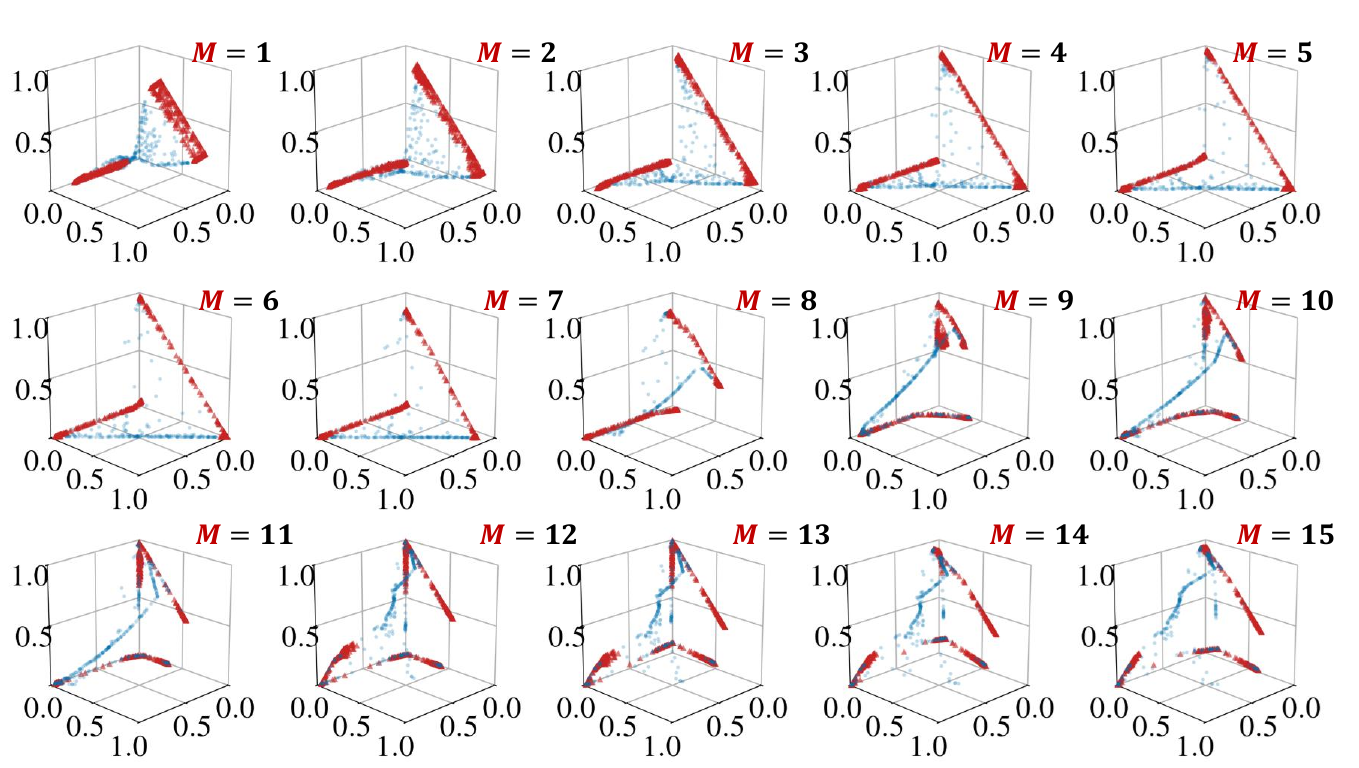}
    \end{center}
    \caption{
    {\bf Dynamical evolution of the intermediate measurement results from $M=1$ to $M=15$.}
    In this Fig, we only present the evolution of Werner states, where the red triangles represent entangled Werner states, and blue circles represent separable ones.
    }
    \label{figs_evo_w}
\end{figure*}

\begin{figure*}[t]
    \begin{center}
        \includegraphics[width=0.98\textwidth]{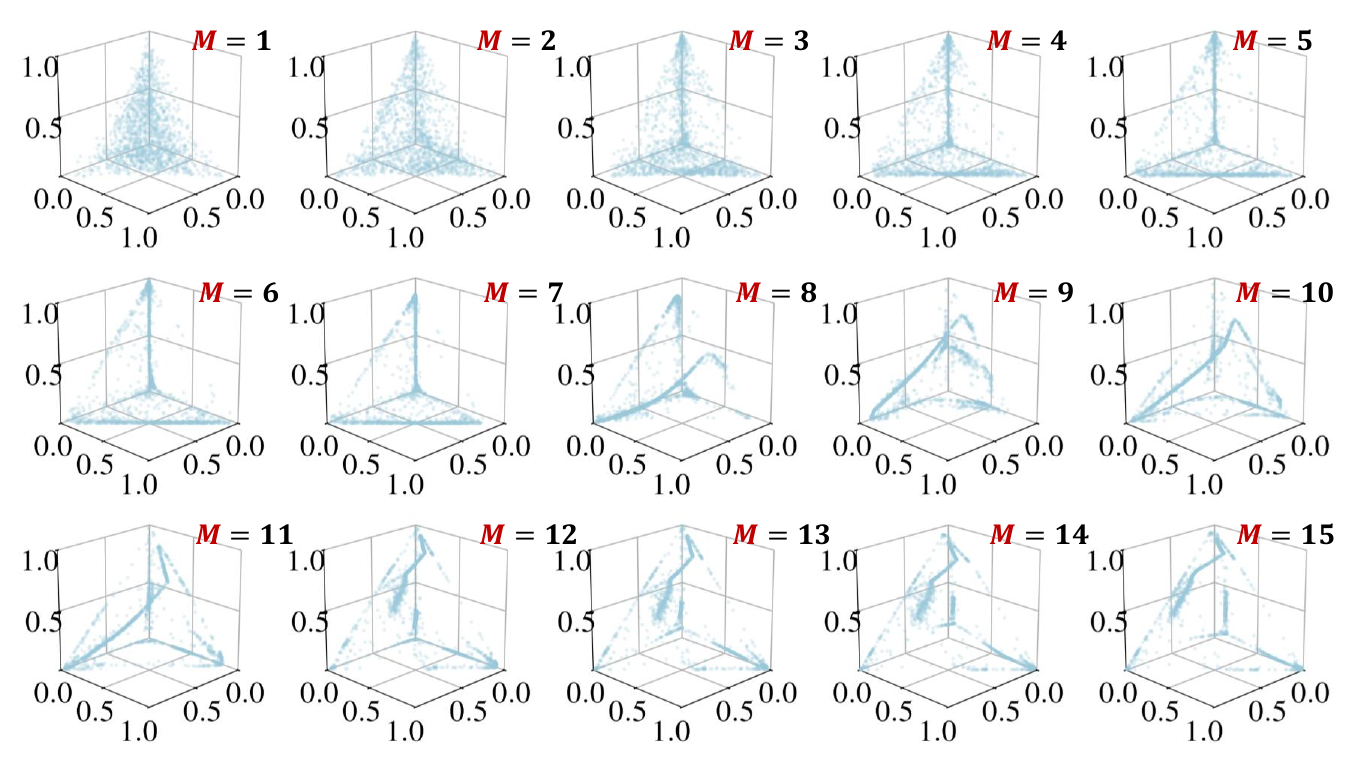}
    \end{center}
    \caption{
    {\bf Dynamical evolution of the intermediate measurement results from $M=1$ to $M=15$.}
    In this Fig, we only present the evolution of random separable states, including both pure and mixed states.
    }
    \label{figs_evo_mix}
\end{figure*}

\textbf{
\begin{figure*}[t]
    \begin{center}
        \includegraphics[width=0.98\textwidth]{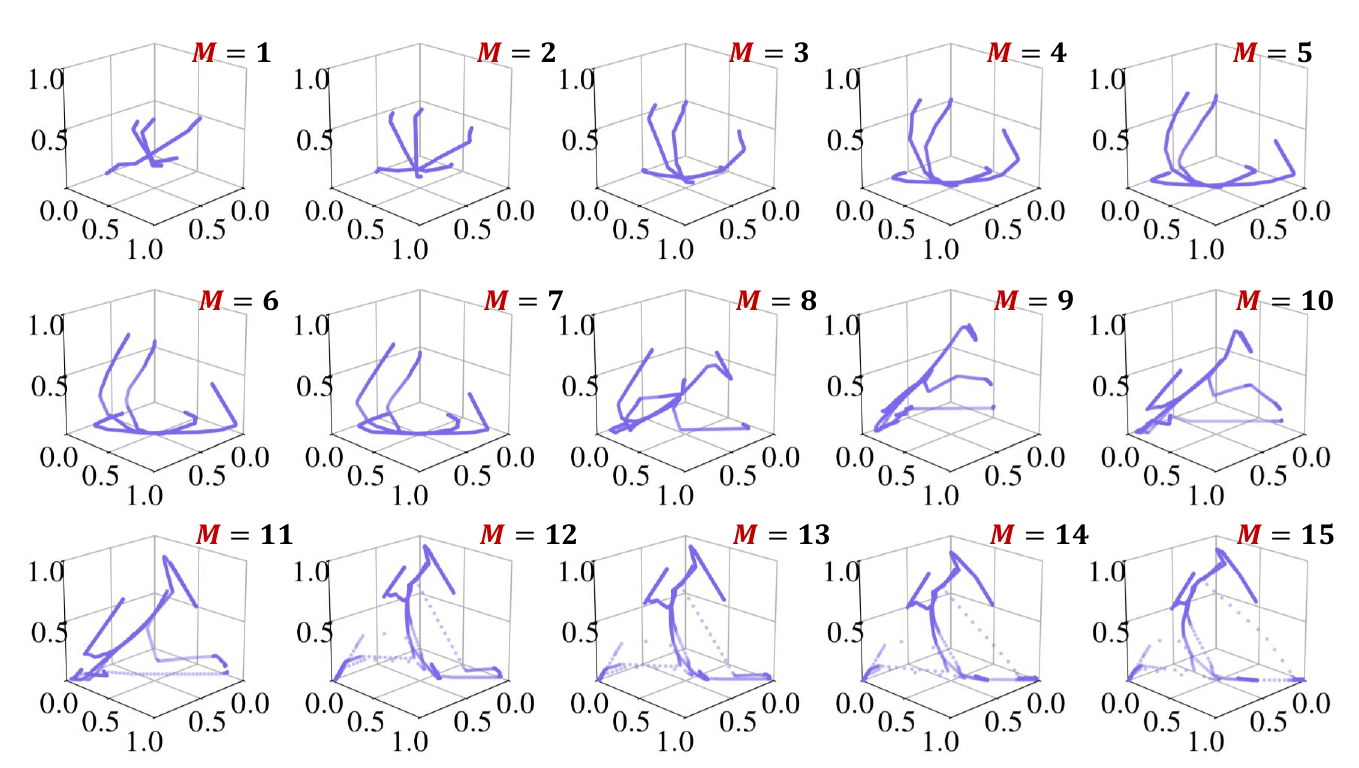}
    \end{center}
    \caption{
    {\bf Dynamical evolution of the intermediate measurement results from $M=1$ to $M=15$.}
    In this Fig, we only present the evolution of adversarial states.
    }
    \label{figs_evo_b}
\end{figure*}}

The architecture of HQRN for entanglement classification is presented in Fig.(3a).
To natively process two-qubit density matrices, each QRB is parameterized to simulate the functional dynamics of a classical residual layer possessing both positive and negative weight components.
For the $k$-th block, we define two independent parameterized quantum circuits, $U^{(k)_\pm}$.
Each unitary operation is constructed using a 15-parameter ansatz spanning two qubits, corresponding to the KAK decomposition. 
The input density matrix $\rho_k$ is evolved through both unitary operations independently. 
A projective measurement in the computational basis, yields two corresponding 4-dimensional probability vectors $p_\pm^{(k)}$.
To introduce the non-linearity, we apply a normalized ReLU activation
\begin{equation}
    F(z_n) = \frac{ReLU(z_n)}{\sum_jReLU(z_j)}
\end{equation}
Thus, we obatin the coefficients $h^{(k)}$
\begin{equation}
    h^{(k)}_n = F\left(\gamma^{(k)}(p^{(k)}_{+,n}-p^{(k)}_{-,n})+b^{(k)}_n\right)
    \label{eq_hk}
\end{equation}
where $\gamma^{(k)}$ and $b^{(k)}_n$ are the scaling factor and bias of the $k\text{-th}$ QRB.
Finally, we mix the intermediate state with the input according to Eq.(\ref{eqs_resc}), by which implements the residual connection.
In this work, we set the residual weighting parameter as $\alpha = 0.5$.

The objective function is designed to effectively separate the manifolds of entangled and separable mixed states within the probability simplex. 
We employ a custom distance-based loss function composed of three terms evaluated over the pairwise Euclidean distance matrix $D_{ij} = \|\mathbf{p}_i - \mathbf{p}_j\|_2$, where $\mathbf{p}_j$ is the 4-d vector obtained by measure the output state $\rho^{(M)}_j$ after $M$ layer of QRBs under computational basis, the subscript $j$ indicates that $\rho^{(M)}_j$ corresponds to the $j$-th training state.
The optimization landscape is defined by the pairwise Euclidean distance matrix $D_{ij} = \|\mathbf{p}_i - \mathbf{p}_j\|_2$.
To ensure the architecture effectively "unfolds" the entangled manifold from the separable background, we employ a contrastive loss function designed to maximize inter-class variance while minimizing intra-class dispersion.
To maintain numerical stability and bound the gradient updates, we apply a hyperbolic tangent scaling to the margin between nearest-neighbor distances:
\begin{equation}
\mathcal{L}_1 = -\frac{1}{N_d} \sum_{i=1}^{N_d} \tanh \left[ \delta_i^{(\text{diff})} - \delta_i^{(\text{same})} \right]
\end{equation}
where $\delta_i^{(\text{diff})} = \min_{j: y_j \neq y_i} D_{ij}$ represents the distance to the nearest state with a disparate label, and $\delta_i^{(\text{same})} = \min_{j: y_j = y_i, j \neq i} D_{ij}$ denotes the distance to the nearest state within the same class.
By this mean, we force the HQRN to navigate the non-diagonal Hilbert space such that the resulting measurement distributions $\mathbf{p}$ cluster into linearly separable regions of the simplex, even when the initial marginal statistics are near-identical.
 
To prevent the collapse of representations from distinct classes, we further introduce an exponential repulsion term active primarily below a critical distance threshold $d_{min}$\begin{equation}
    \mathcal{L}_{2} = \langle \exp(-D_{\text{diff}} / d_{min}) \rangle
\end{equation}
Additionally, for states that achieve an identical mapping ($D_{ij} < 10^{-2}$ for $y_i \neq y_j$), a static, large scalar penalty $\mathcal{L}_{3}$ is applied to enforce strict manifold unfolding.
Here we set $\mathcal{L}_{3}\sim 100N_{close}$, where $N_{close}$ is the number of pairs that too close to each other, $D_{ij} < 10^{-2}$.
The total loss minimized at each layer is 
\begin{equation}
    \mathcal{L}_q = \mathcal{L}_{1} + \beta \mathcal{L}_{2} + \mathcal{L}_{3}
\end{equation}
in the numerical simulation, we set $\beta = 5$.

The HQRN is trained by utilizing a greedy, layer-wise protocol.
The network is constructed by sequentially optimizing one QRB at a time. For the $k$-th block, the parameter set is optimized while keeping the parameters of all preceding blocks $\{1, \dots, k-1\}$ frozen.
To mitigate trapping in local minima on the highly non-convex loss landscape, each layer undergoes $N_{iter} = 5$ independent random initializations. 
The network executes a maximum of 200 optimization steps per initialization, selecting the parameter set $\Theta_k^*$ that achieves the lowest total loss before appending the finalized block to the architecture and propagating the dataset forward to train the next layer.

These $M$ QRBs are optimized to maximize the distinguishability of the measurement results.
The output state of the $M$-th QRB $\rho^{(M)}$ is projected into a probability distribution vector by applying measurement on computational basis, denoted as a 4D probability vector $(p_{00}, p_{01}, p_{10}, p_{11})$, which is depicted within an $\mathbb{R}^3$ subspace $(p_{00}, p_{01}, p_{10})$, as the fourth component can be obtained by $p_{11} = 1 - p_{00} - p_{01} - p_{10}$.
Since then, the input quantum states are transformed to 4-d probability vectors, which can be mapped as diagonal density matrices.
Therefore, the following 10 QRBs are equivalent to the classical counterpart, and the recursive transformation is given by Eq.(\ref{eqs_update_y}).
Similarly to the optimization in digit recognition part, here we optimize the CRBs to obtain the weights $W$ and bias $b$, then reconstruct the QRBs via unitary dilation and Trotter decomposition.

In the optimization process of these 10 CRBs, we apply a first-order gradient-based approach.
The models are optimized using binary cross entropy.
As the dataset exhibits an inherent asymmetry between the number of separable states ($y=0$) and entangled states ($y=1$),
to prevent the decision boundary from skewing toward the majority class, a positive weight scalar $w_p = N_{sep} / N_{ent}$ is integrated into the loss function, yielding
\begin{equation}
    \mathcal{L}_c = - \frac{1}{N_d} \sum_{i=1}^{N_d} \Big[ w_p y_i \ln(\hat{y}_i) + (1 - y_i) \log(1 - \hat{y}_i) \Big]
\end{equation}
The weight $w_p = N_{sep} / N_{ent}$ forces the gradient to pay more attention to the minority class (the entangled states), ensuring the decision boundary is physically meaningful rather than just statistically biased.
In numerical simulation, we apply the Adam optimizer with a learning rate of $10^{-3}$, providing adaptive moment estimation to navigate the constrained parameter space of the simplex-bounded blocks.

By the end, we present the Layer-wise dynamical evolution of the intermediate measurement distributions for Werner states, random mixed states, and the adversarial states in Fig.(\ref{figs_evo_w}), Fig.(\ref{figs_evo_mix}) and Fig.(\ref{figs_evo_b}), respectively.
The trajectory of two-qubit Werner states is visualized within the 3-simplex subspace. 

\end{document}